\documentclass[letterpaper]{article} 
\usepackage{aaai25}  
\usepackage{times}  
\usepackage{helvet}  
\usepackage{courier}  
\usepackage[hyphens]{url}  
\usepackage{graphicx} 
\usepackage{multirow}
\usepackage{natbib}  
\usepackage{caption} 
\usepackage{algorithm}
\usepackage{algorithmic}

\usepackage{xcolor}

\usepackage{newfloat}
\usepackage{listings}
\DeclareCaptionStyle{ruled}{labelfont=normalfont,labelsep=colon,strut=off} 
\floatstyle{ruled}
\newfloat{listing}{tb}{lst}{}
\floatname{listing}{Listing}
\usepackage{booktabs}
\usepackage{tcolorbox}
\tcbuselibrary{listings}
\usepackage{tikz}
\usetikzlibrary{arrows.meta,positioning}

\newtcblisting{promptbox}[1]{%
  title=#1,
  listing only,
  colback=white!95!gray,
  colframe=gray,
  width=\linewidth,
  arc=0mm,
  boxrule=0.5mm,
  colbacktitle=white!95!gray,
  coltitle=black,
  fonttitle=\bfseries,
  left=6pt,
  right=6pt,
  top=6pt,
  bottom=6pt,
  listing options={basicstyle=\small\ttfamily, breaklines=true}
}

\title{Podcasts as a Medium for Participation in Collective Action:\\ A Case Study of Black Lives Matter}

\author {
    Theodora Moldovan\textsuperscript{\rm 1,*},
    Arianna Pera\textsuperscript{\rm 2},
    Davide Vega\textsuperscript{\rm 1},
    Luca Maria Aiello\textsuperscript{\rm 2,3}
}
\affiliations {
    \textsuperscript{\rm 1}InfoLab, Department of Information Technology, Uppsala University\\
    \textsuperscript{\rm 2}IT University of Copenhagen\\
    \textsuperscript{\rm 3}Pioneer Centre for AI\\
    $^*$Corresponding author. Email: theodora.moldovan@nek.uu.se
}

\begin{document}

\maketitle

\begin{abstract}
We study how participation in collective action is articulated in podcast discussions, using the Black Lives Matter (BLM) movement as a case study. While research on collective action discourse has primarily focused on text-based content, this study takes a first step toward analyzing audio formats by using podcast transcripts. Using the Structured Podcast Research Corpus (SPoRC), we investigated spoken language expressions of participation in collective action, categorized as \textit{problem-solution}, \textit{call-to-action}, \textit{intention}, and \textit{execution}. We identified podcast episodes discussing racial justice after important BLM-related events in May and June of 2020, and extracted participatory statements using a layered framework adapted from prior work on social media. We examined the emotional dimensions of these statements, detecting eight key emotions and their association with varying stages of activism. We found that emotional profiles vary by stage, with different positive emotions standing out during \textit{calls-to-action}, \textit{intention}, and \textit{execution}. We detected negative associations between collective action and negative emotions, contrary to theoretical expectations. Our work contributes to a better understanding of how activism is expressed in spoken digital discourse and how emotional framing may depend on the format of the discussion.
\end{abstract}

\begin{links}
    \link{Code}{https://github.com/todomoldovan/blm_in_podcasts}
\end{links}

\section{Introduction} 
The pressing social and global challenges of our time, such as climate change and marginalization, demand collective, grassroots efforts to drive meaningful change~\citep{lteif_climate_2024}. For these efforts to be effective, digital communication channels play a role in disseminating messages widely and rapidly. Such platforms do more than spread information: they shape narratives within collectives, reveal how individuals and groups express intent, support, or commitment to shared causes~\citep{margetts_political_2016, chiovaro_online_2021}, and serve as crucial tools for movements to organize and mobilize supporters, amplify their reach, articulate demands, and influence public discourse. Understanding these communicative dynamics is therefore essential for sustaining and advancing democratic processes~\citep{margetts_political_2016,lucchini_reddit_2022}.

At the heart of these efforts lies collective action: the shared expressions of participation, commitment, and resistance that allow movements to build momentum and legitimacy. Recent advances in natural language processing make it possible to systematically trace such expressions in online texts~\citep{pera_extracting_2025}, opening new opportunities to study the role of digital communication in sustaining collective action. Emotions play a critical role in this collective process. They motivate participation by fostering solidarity among allies and outrage toward perceived opponents, while also shaping the perceptions of those not directly involved in the movement~\citep{van_troost_emotions_2013, jasper_emotions_2018}. Examining how emotions co-occur with expressions of collective action is therefore key to unpacking the mechanisms through which digital mobilization unfolds.

Podcasts have emerged as influential venues for shaping public opinion, mobilizing audiences, and driving political engagement~\citep{aufderheide_podcasting_2020, bottomley_podcasting_2015, pew_research_center_audio_2023}. As part of a broader shift towards immersive formats~\citep{pew_research_center_digital_2023}, they represent a growing arena for digital discourse. Despite their reach, podcasts remain underexplored in studies of social movements and activism. At the same time, they provide a uniquely rich communicative environment, where language, voice, and tone interact to convey meaning. This makes them an important setting for studying how affective digital communication supports social mobilization. 

This paper focuses on racial justice, using the Black Lives Matter (BLM) movement as a prominent example of a digitally mediated social movement. BLM’s online presence can be traced back to 2013, when its founders started using the \#BlackLivesMatter hashtag on Twitter (now X) in response to the acquittal of George Zimmerman in the fatal shooting of Trayvon Martin. Later in 2014, the hashtag served as a digital catalyst for mobilization after the shooting of Michael Brown and the ensuing Ferguson, Missouri protests, marking the beginning of a social movement that leveraged online platforms for mobilization towards collective action~\cite{freelon_beyond_2016, mundt_scaling_2018}. The death of George Floyd in 2020 sparked nationwide protests led by BLM, igniting widespread discussions on police brutality, systemic racism, and related cases like those of Breonna Taylor and Ahmaud Arbery~\cite{anderson_blacklivesmatter_2020, nguyen_progress_2021}. 

In this paper, we draw on the Structured Podcast Research Corpus~\citep{litterer_mapping_2025}, which contains over one million podcast transcripts from May and June 2020 — a period of heightened racial justice activism following the murder of George Floyd. We use a racial justice classifier to identify episodes discussing racial justice and analyze how expressions of collective action emerge in those conversations. Finally, we examine how collective action expressions relate to emotions commonly associated with social movement mobilization contexts. Our analysis is guided by the following research questions:

\vspace{4pt} \noindent \textbf{RQ1.} 
\textit{How do podcasts reflect shifts in racial justice discourse in response to the intensity of events?}

\vspace{2pt} \noindent \textbf{RQ2.} \textit{To what extent do podcasts serve as mediums where participation in collective action around racial justice is articulated and discussed?}

\vspace{2pt} \noindent \textbf{RQ3.} \textit{How are emotions expressed in relation to different levels of collective action in racial justice discourse within podcasts?}

\vspace{4pt} Our contributions are twofold: (1) We expand the study of digital activism beyond short-form text platforms (e.g., X) by introducing podcasts as a large-scale, audio-first medium for analyzing social movements. Using over 150k episode transcripts, we show how offline racial justice events shaped the emergence of collective action in podcast discussions during the BLM protests of 2020.
(2) We advance theories of collective action and emotions by analyzing their co-occurrence, revealing how different emotional registers (e.g., \textit{optimism}, \textit{anger}) are intertwined with varying levels of mobilization.
Together, these contributions reveal how voice-based media broaden our understanding of how collective action is expressed, sustained, and emotionally charged in the digital age.

\section{Methods}

\begin{figure}[t!]
\centering
\resizebox{0.9\columnwidth}{!}{
\begin{tikzpicture}[
  font=\normalsize,
  node distance=9mm,
  >=Latex,
  block/.style={draw, rounded corners=2pt, align=center, thick, fill=gray!10, inner sep=4pt, text width=0.8\columnwidth},
  classifier/.style={draw, dashed, rounded corners=2pt, align=center, fill=blue!5, inner sep=3pt, text width=0.8\columnwidth},
  arrow/.style={-Latex, thick}
]

\node (start)   [block] {\textbf{155,784} episodes with racial justice mentions\\(keyword search)};
\node (clf1)    [classifier, below=of start] {Apply binary collective action classifier\\(on all sentences)};
\node (clf2)    [classifier, below=of clf1] {Apply multi-class collective action classifier\\(on collective action sentences)};
\node (keep1)   [block, below=of clf2] {\textbf{152,539} episodes with $\geq$1 collective action statement};
\node (clf3)    [classifier, below=of keep1] {Apply racial justice classifier\\(on collective action sentences)};
\node (keep2)   [block, below=of clf3] {\textbf{139,200} episodes with collective action statements about race};
\node (clf4)    [classifier, below=of keep2] {Apply emotion classifiers\\(on all sentences)};

\draw[arrow] (start) -- (clf1);
\draw[arrow] (clf1) -- (clf2);
\draw[arrow] (clf2) -- (keep1);
\draw[arrow] (keep1) -- (clf3);
\draw[arrow] (clf3) -- (keep2);
\draw[arrow] (keep2) -- (clf4);

\end{tikzpicture}
}
\caption{Filtering and data processing pipeline from keyword-matched episodes to the final sample. Dashed boxes refer to steps in which classifiers are applied.}
\label{fig:pipeline}
\end{figure}
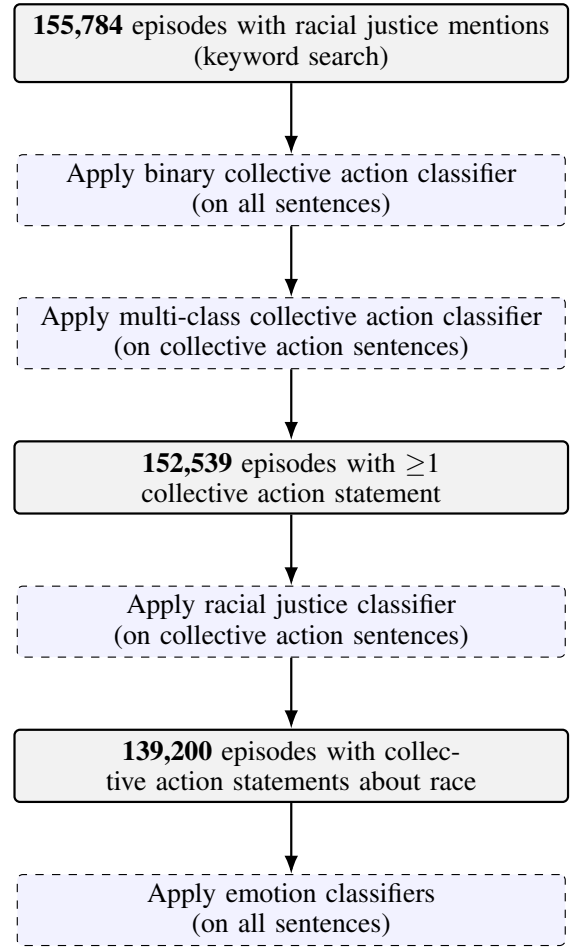

Starting from a large corpus of English-language podcasts, we progressively filtered our dataset to episodes that both mention racial justice and contain explicit statements of participation in collective action (see Figure~\ref{fig:pipeline}). At each stage, specialized classifiers were applied to either identify statements that are about racial justice and collective action, or capture the emotional valence of participation in collective action. In this section we describe the data, preprocessing steps, classification models, and statistical analyses performed in our study. Classifier validation is described separately in the ``Validation'' section, which follows the results. 

\subsection{Data}

The Structured Podcast Research Corpus (SPoRC)~\cite{litterer_mapping_2025} consists of English language podcast episodes from May and June of 2020, and contains over 1.1M transcripts of episodes identified through the Podcast Index\footnote{\url{https://podcastindex.org/}}, a public database with information on over 4 million different shows. The episodes were collected in audio format via RSS feeds and automatically transcribed using \verb|whisper|~\citep{radford_robust_2022}. The transcripts also include non-speech tags such as ``(laughing)'', which we deliberately retain, as these cues can be integral to conveying emotion and context in spoken language. 
We further preprocessed the episode-level transcripts by splitting the text into segments using punctuation markers. We applied all classification models at the sentence level.

\subsection{Keyword Search}

We began with $155,784$ podcast episodes containing at least one mention of racial justice, identified through a transcript-level keyword search. See Table~\ref{tab:keywords} in the Appendix for the full list of keywords used to filter episodes.
        
\subsection{Participation in Collective Action Classification}
\label{subsec:class}

We adopted the approach by~\citet{pera_extracting_2025} and thus defined collective action as \textit{``any online or offline effort that can be taken to mitigate a collective action problem.''}

We used the layered learning framework developed in the work, which is grounded in the theoretical framework of social movement mobilization and structured as a two-stage classification pipeline. In the first stage, we use a \verb|RoBERTa|-based classification model\footnote{\url{https://huggingface.co/ariannap22/collectiveaction_roberta_simplified_synthetic_weights}} to assess whether a particular piece of text indicates participation in collective action or not. If participation is identified, in the second stage we use a fine-tuned Llama3 model\footnote{\url{https://huggingface.co/ariannap22/collectiveaction_sft_annotated_only_v6_prompt_v6_p100_synthetic_balanced_more_layered}} in a multi-class classification setting to determine the specific level of participation expressed.
This can be one of the following four:
\begin{enumerate}
    \item \textbf{Problem-solution}: Identifying a collective action issue, assigning blame, or proposing solutions.
    \item \textbf{Call-to-action}: Urging others to join or support a cause.
    \item \textbf{Intention}: Expressing willingness or interest in taking action.
    \item \textbf{Execution}: Reporting involvement in collective action initiatives.
\end{enumerate}

We applied the binary classifier to all sentences to identify candidate statements of participation in collective action. On the sentences flagged positively, we applied the multi-class classifier for finer-grained labeling. Episodes without any statement expressing participation in collective action were discarded, leaving $152,539$ episodes.

\subsection{Racial Justice Classification}

Since the initial sample was constructed through keyword matching, not all collective action statements explicitly addressed racial justice. To evaluate whether sentences expressing participation in collective action specifically concerned issues of racial justice, we employed a zero-shot classification approach using the \texttt{Llama-3.3-70B-Instruct} model\footnote{\url{https://huggingface.co/unsloth/Llama-3.3-70B-Instruct-bnb-4bit}}~\cite{grattafiori_llama_2024_short}.

We applied this classifier exclusively to sentences that had already been identified as expressions of collective action, thereby refining our analysis to focus on the social movement of interest. The model was prompted to determine whether a given sentence pertained to racial justice or not. The full prompt used for this task is provided in Appendix~\ref{app:prompts}.

This step produced a final sample of $139,200$ episodes containing statements that expressed participation in collective action related to racial justice.

\subsection{Emotion Classification}

Finally, we focused on emotions expressed in sentences within the retained episodes. To guide our analysis, we relied on the GoEmotions taxonomy~\cite{demszky_goemotions_2020}, which offers a comprehensive set of 27 categories spanning negative, positive and ambiguous sentiments. From this taxonomy, we selected emotions according to two criteria: (i) their relevance in the literature on emotions in collective actions, and (ii) their empirical reliability, defined as a class-wise FI score of at least $0.3$. This ensured that our analysis remained both theoretically grounded and methodologically robust, allowing for a meaningful comparison with previous studies focused on emotions in the Black Lives Matter movement.

We used a model\footnote{\url{https://huggingface.co/SamLowe/roberta-base-go_emotions}} trained from \verb|RoBERTa-base| on the GoEmotions dataset~\cite{demszky_goemotions_2020} for multi-label classification. 
We selected eight emotions based on literature relevance and empirical reliability and report them in Table~\ref{tab:emotions}, together with their F1 scores based on the model applied to the GoEmotions dataset.
The full list of emotions in the dataset and the F1 metrics are reported in Table~\ref{tab:big_emotions} in the Appendix.

\setlength{\tabcolsep}{4pt} 
\begin{table*}[t]
\centering
\begin{tabular}{@{}lp{2.2cm}p{7cm}cc@{}}
\toprule
\textbf{Polarity} & \textbf{Emotion} & \textbf{Definition} & \textbf{F1} & \textbf{\% of collective action participation} \\ 
\midrule
\parbox[t]{2mm}{\multirow{5}{*}{\rotatebox[origin=c]{90}{Negative}}} 
  & \textbf{Anger}   & A strong feeling of displeasure or antagonism. & 0.479  & 0.851 \\
  & \textbf{Disgust} & Revulsion or strong disapproval aroused by something unpleasant or offensive. & 0.453  & 0.067 \\
  & \textbf{Fear}    & Being afraid or worried. & 0.671  & 0.392 \\
  & \textbf{Sadness} & Emotional pain, sorrow. & 0.550  & 1.155 \\
\midrule
\parbox[t]{2mm}{\multirow{5}{*}{\rotatebox[origin=c]{90}{Positive}}} 
  & \textbf{Caring}  & Displaying kindness and concern for others. & 0.372  & 0.052 \\
  & \textbf{Joy}     & A feeling of pleasure and happiness. & 0.600  & 0.313 \\
  & \textbf{Love}    & A strong positive emotion of regard and affection. & 0.802  & 0.841 \\
  & \textbf{Optimism}& Hopefulness and confidence about the future or the success of something. & 0.481  & 1.581 \\
\bottomrule
\end{tabular}
\caption{Definitions of selected emotions (grouped by polarity), classifier metrics and percentage of each emotion found in the set of expressions of collective action.}
\label{tab:emotions}
\end{table*}

\subsection{Associations}

To examine the relationship between each emotion and the level of participation in collective action, we computed four separate odds ratios, one for each extracted level of participation in collective action (\texttt{CA}). These odds ratios served as statistical measures of association between the presence of an emotion and each specific level of collective action engagement. We denote the binarized score of a sentence $s$ containing emotion $e$ as $e(s)$.

We computed separate odds ratios for each level of participation instead of a single odds ratio for participation vs. non-participation because the class distribution across participation in collective action levels is imbalanced, with the majority of paragraph turns falling under the \textit{problem–solution} category.

For a given level $\ell$ of collective action participation, let $S_{\texttt{CA}_\ell}$ denote the set of sentences labeled with that level. We computed the conditional probability that a sentence expressing emotion $e$ appears among those associated with level $\ell$, defined as:

\begin{equation} p(e \mid \texttt{CA}_\ell) = \frac{\sum_{s \in S_{\texttt{CA}_\ell}} e(s)}{|S_{\texttt{CA}_\ell}|}, \end{equation}

A similar conditional probability $p(e \mid \overline{\texttt{CA}_\ell})$ was defined for sentences not associated with collective action. Then, the odds ratio between $e$ and $\texttt{CA}_\ell$ was computed as:

\begin{equation}
OR_\ell(p(e \mid \texttt{CA}_\ell), p(e \mid \overline{\texttt{CA}_\ell})) = \frac{p(e \mid \texttt{CA}_\ell) / (1 - p(e \mid \texttt{CA}_\ell))}{p(e \mid \overline{\texttt{CA}_\ell}) / (1 - p(e \mid \overline{\texttt{CA}_\ell}))}.
\end{equation}

Each $OR_\ell$ indicates whether an emotion $e$ is more likely to co-occur with collective action than without. An $OR_\ell > 1$ implies a positive association and $OR_\ell < 1$ implies a negative one. Additionally, when $OR_\ell \approx 1$ (is very close to 1), no definitive conclusions can be drawn, as the likelihood of the outcome is approximately the same in both groups, suggesting no meaningful association.

\section{Results}

Our analysis proceeded in three stages, aligned with our research questions. First, we examined how podcast discussions of racial justice unfolded over time during May and June 2020, contextualizing shifts in volume against key events of the BLM movement. Second, we analyzed the presence of collective action discourse, assessing how podcasts engaged with the movement. Finally, we explored the emotional landscape of these discussions, investigating how different emotions were associated with varying levels of collective action. Together, these results provide a multi-dimensional view of how podcasts engaged with BLM during this period, identifying both mobilization rhetoric and the affective dimensions of collective action.

\subsection{BLM Discussions in Podcasts}

\begin{figure}[t!]
    \centering
    \includegraphics[width=\columnwidth]{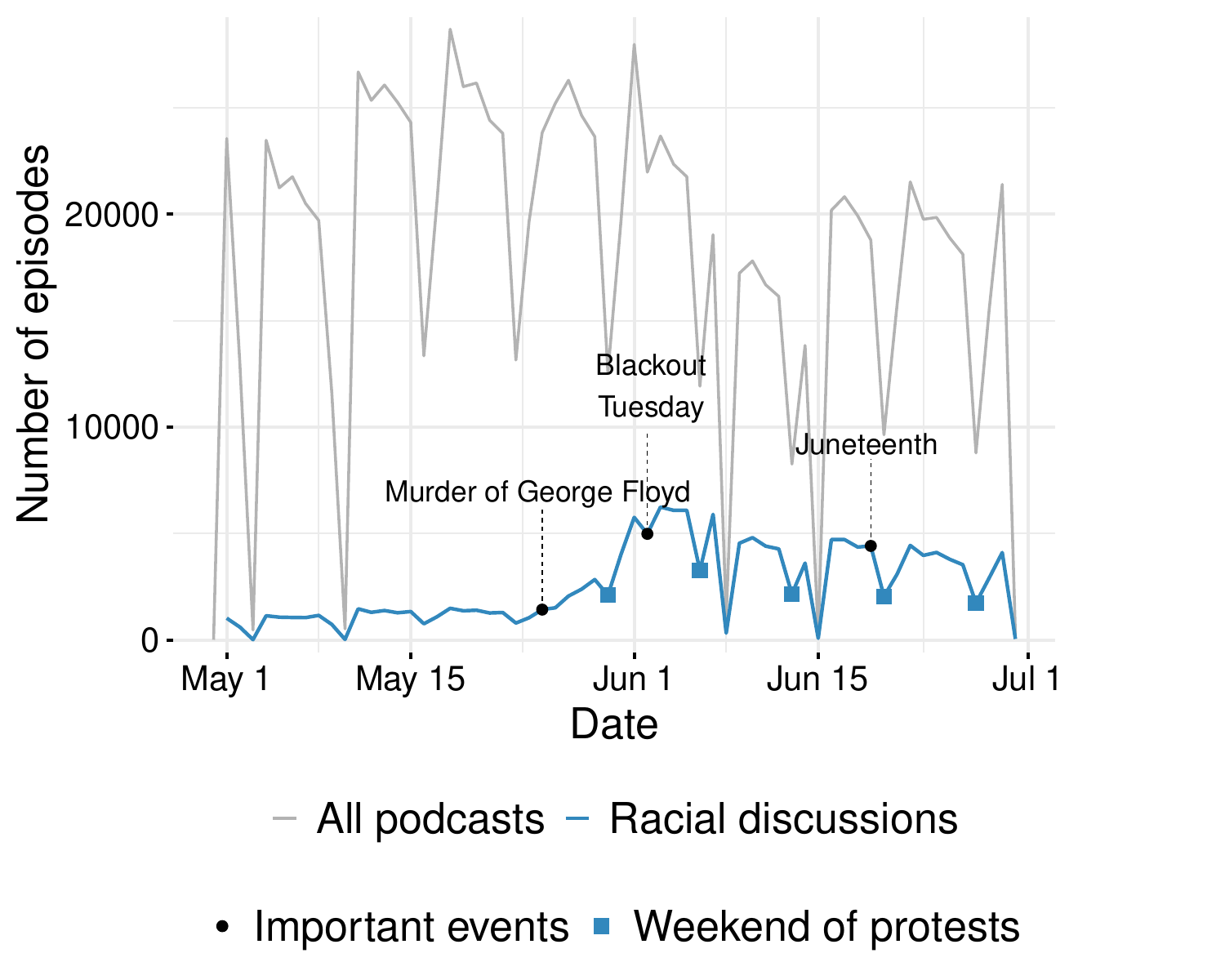}
    \caption{Time series of episodes under analysis.}
    \label{fig:time_series}
\end{figure}

The SPoRC dataset extensively covers all English language podcast episodes distributed through public RSS feeds during the months of May and June of 2020. As shown in Figure~\ref{fig:time_series}, more podcast episodes were released in May than in June. However, the opposite trend is observed for episodes addressing racial issues.

Prior to the murder of George Floyd on May 25, the number of episodes discussing racial issues was relatively low and stable. Immediately following the event, this number increased sharply, followed by an additional surge after the first week of nationwide protests. The volume of race-related content reached its peak shortly after Blackout Tuesday (June 2), a day of action initiated by the music industry in protest of racism and police violence. The event was widely discussed across social media platforms, where millions of users posted black squares with the hashtag \#BlackOutTuesday. Although the campaign was criticized as performative, the debate over its nature itself became a salient theme in podcast coverage, coinciding with the highest level of race-related discussion in the observed period.

From the second and largest weekend of protests through the end of June, the number of discussions declined to a level that was still twice as high as the pre-May 25 baseline. Another notable spike occurred in the days leading up to Juneteenth (June 19), the annual commemoration of the end of slavery in the United States, which was later recognized as a federal holiday in 2021.

An additional pattern is the recurring increase in race-related discussions following the weekends of protests. This trend may reflect podcast release schedules, which often cluster at the beginning of the week, or it could indicate heightened media coverage and public attention following weekends of protest activity.

\subsection{Participation in Collective Action in Podcasts}

We found that $89.34\%$ of podcasts on the topic of racial justice contained at least one sentence expressing participation in collective action. An additional $8.56\%$ of the episodes discussing racial justice contained statements of participation in collective action unrelated to race. Since podcast episodes often cover multiple topics, these remaining expressions may reflect other movements or social issues, such as the COVID-19 pandemic, which was another dominant theme of public discourse at the time.

Expressions of collective action appeared across a wide range of types of podcasts (see Table~\ref{tab:categories} in the Appendix for a full list of categories). Over half of the episodes that address racial themes belong to podcasts categorized under religion, society, or news. While the high frequency of religious podcasts may seem unexpected, it reflects the fact that religion was the single largest category of all episodes published during this period, consisting largely of pre-recorded sermons~\cite{litterer_mapping_2025}. Prior research has shown that in the months following the summer 2020 protests, congregants who heard sermons addressing race and policing were more likely to support BLM and participate in racial justice efforts~\cite{brown_race_2023}.

Although the majority of podcasts touching on racial justice contained some form of collective action discourse, these discussions were generally brief: in $80\%$ of the episodes, the number of direct expressions of participation did not exceed $25$, despite episodes averaging 564 sentences.

\begin{figure}[t]
    \centering
    \includegraphics[width=\columnwidth]{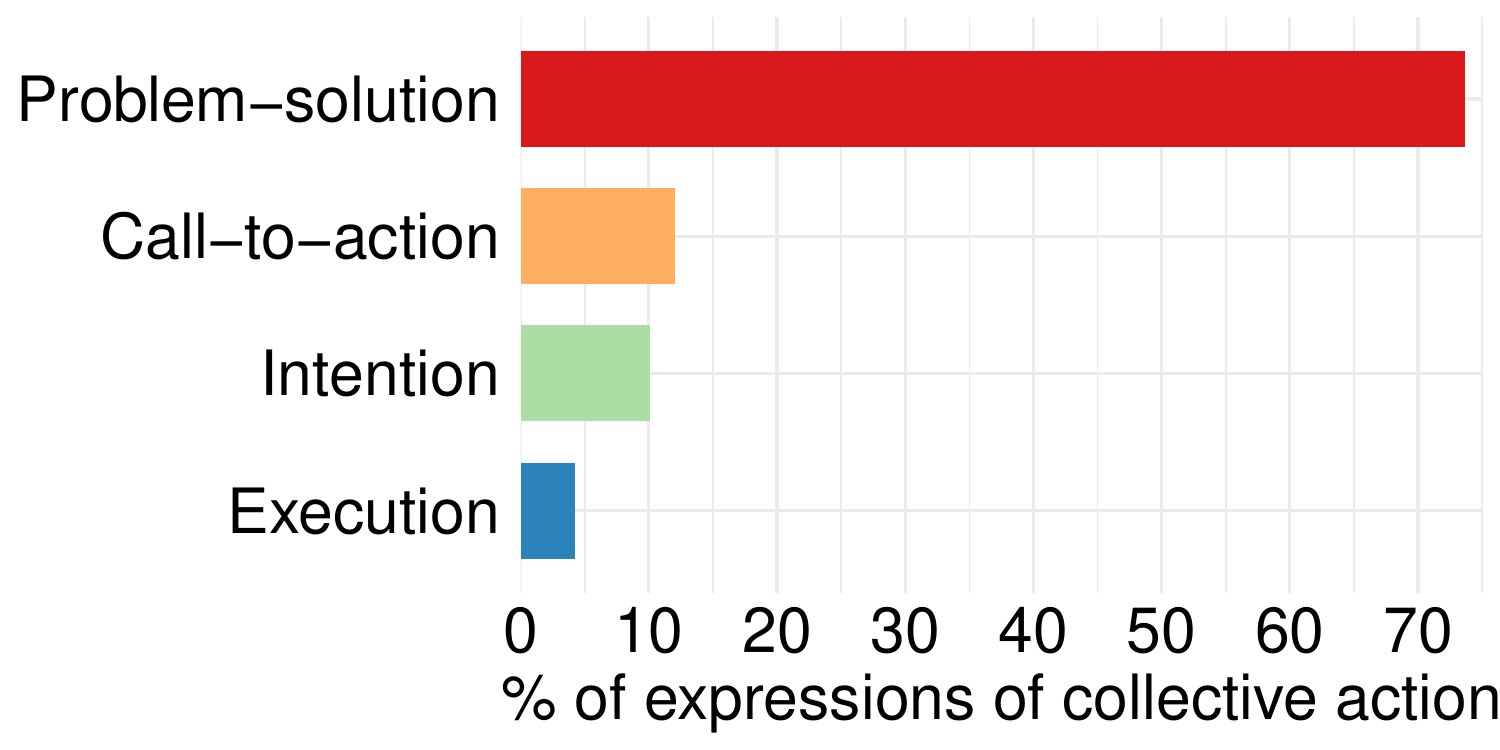}
    \caption{Percentage of each level of participation out of all collective action sentences.}
    \label{fig:class_percent}
\end{figure}

Within episodes, most expressions of collective action belonged to the \textit{problem-solution} class, which accounted for an average of $73.63\%$ of all statements (Figure~\ref{fig:class_percent}). This category, which includes identifying collective patters, assigning blame, and proposing solutions, has also been shown to dominate discussions in other contexts, such as climate-related subreddits~\cite{pera_extracting_2025}. The remaining classes were more evenly distributed: \textit{call-to-action} ($12.02\%$), \textit{intention} ($10.08\%$), and \textit{execution} ($4.27\%$). At the episode level, nearly all episodes contained at least one \textit{problem-solution} statement ($94.43\%$), and more than half also included \textit{call-to-action} ($55.74\%$) and \textit{intention} ($54.12\%$) statements, while fewer included explicit \textit{execution} statements ($33.44\%$).

These figures suggest that while discussion of racial justice is somewhat frequent, explicit reports of participation in action events are more limited, aligning with our expectations and the broader focus on reflection and critique found in podcast conversations.

\subsection{Emotions of Participation in BLM}

To relate participation in collective action with emotions, we detected eight emotions that prior literature \cite{field_analysis_2022, chevrette_humanizing_2025, ellefsen_black_2022, cappelli_black_2020, vance_black_2022} has suggested being related to the BLM discourse: \textit{anger}, \textit{disgust}, \textit{fear}, \textit{sadness}, \textit{joy}, \textit{optimism}, \textit{love}, and \textit{caring}. We applied the \verb|RoBERTa-base| model fine-tuned on the GoEmotions dataset to detect each emotion in all sentences in our dataset, finding that $6.26\%$ of sentences co-occur with at least one emotion. The proportion of sentences with at least one emotion detected does not vary much between the expressions of collective action and the remaining sentences. 

\textit{Optimism} is the most common emotion found within expressions of participation in collective action on race, followed by \textit{sadness}, while \textit{caring} and \textit{disgust} appear the least (Table~\ref{tab:emotions}). The share of sentences in the full dataset containing each emotion is provided in Table~\ref{tab:emodist}. We emphasize that a single sentence can convey more than one emotion.

\begin{figure*}[t]
    \centering
    \includegraphics[width=1.9\columnwidth]{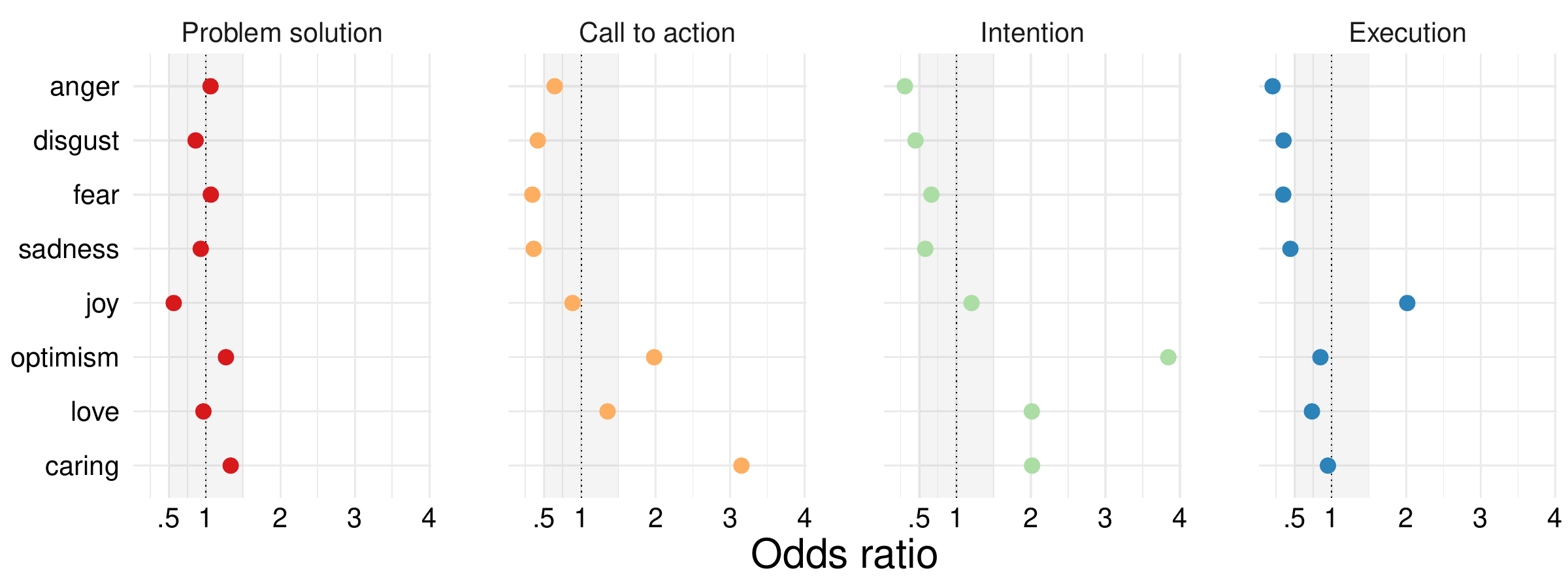}
    \caption{Odds ratios for the presence of emotions (rows) in sentences containing each level of collective action participation (columns), compared to sentences without that level. The vertical dashed line represents odds ratio $= 1$ and the grey area covers the $[0.5, 1.5]$ interval. An odds ratio $> 1$ implies a positive association and $< 1$ implies a negative one.}
    \label{fig:lolliplot_emotions}
\end{figure*}

We computed odds ratios for each level of collective action participation and each emotion to determine whether certain emotions are more likely to co-occur with certain levels of participation (Figure~\ref{fig:lolliplot_emotions}). All odds ratios reported are statistically significant, and none of the emotional categories presented are highly unbalanced. Confidence intervals are reported in Table~\ref{tab:oddsdeets}. We do not discuss results within the $[0.5, 1.5]$ interval as the odds ratios are too close to $1$ to reach a meaningful conclusion.

We do not find any strong association between any of the emotions and \textit{problem-solution} statements. For \textit{call-to-action} statements, we find that negative emotions (\textit{disgust}, \textit{fear}, \textit{sadness}) have negative associations and positive emotions (\textit{optimism}, \textit{caring}) have positive associations. The same pattern for positive emotions is even more evident for \textit{intention}. Small, negative associations exist with \textit{anger} and \textit{disgust}, while \textit{optimism}, \textit{love} and \textit{caring} show high positive associations. In the \textit{execution} stage, however, only \textit{joy} is positively associated with such statements, while all negative emotions (\textit{anger}, \textit{disgust}, \textit{fear}, \textit{sadness}) are negatively related.

These patterns resonate with the functional role of different stages of collective action. \textit{Calls-to-action}, which appeal to solidarity and mutual support, are positively associated with emotions of \textit{caring}. \textit{Intention} statements, inherently future-oriented, are aligned with \textit{optimism}, an emotion tied to anticipation. Finally, \textit{execution} is positively associated with \textit{joy}, reflecting the affective dimension of accomplishment. While we cannot claim causal links, these associations highlight a meaningful correspondence between the emotional tone of discourse and the rhetorical purpose of different participation stages.

\begin{figure}[t]
    \centering
    \includegraphics[width=0.9\columnwidth]{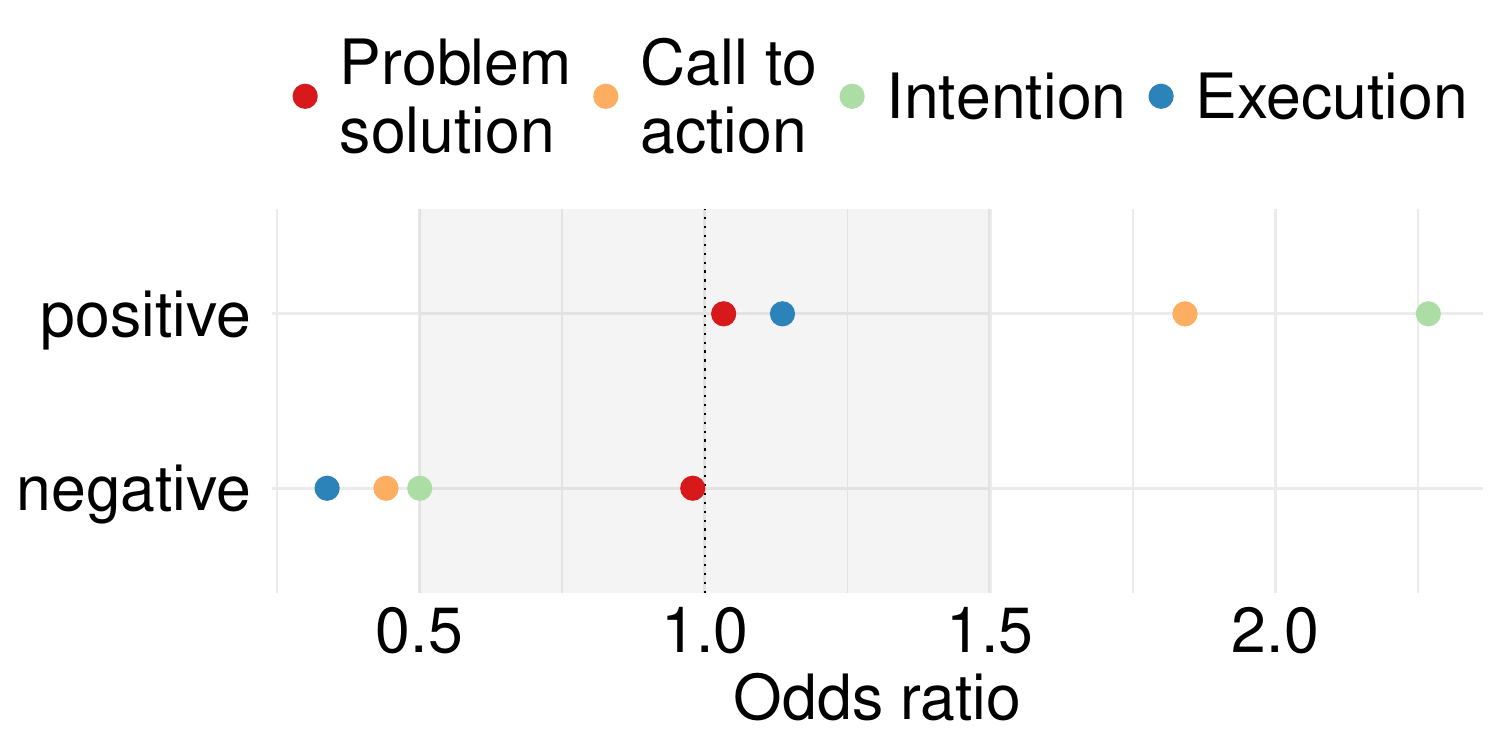}
    \caption{Odds ratios for the presence of a sentiment-based group of emotions (rows) in sentences containing each level of collective action participation (columns), compared to sentences without that level. The vertical dashed line represents odds ratio $= 1$ and the grey area covers the $[0.5, 1.5]$ interval. An odds ratio $> 1$ implies a positive association and $< 1$ implies a negative one.}
    \label{fig:lolliplot_sentiment}
\end{figure}

A common problem identified in previous studies~\cite{demszky_goemotions_2020, field_analysis_2022} is the confusion of similar emotions (e.g., \textit{love} vs. \textit{caring}), both in humans and machines. As we notice similar patterns in related emotions (e.g., \textit{anger} and \textit{disgust}), we group the emotions in two sentiment-based categories informed by prior work~\cite{demszky_goemotions_2020}. Each group comprises four emotions and we repeat the analysis for each sentiment grouping and each level of participation in collective action (Figure~\ref{fig:lolliplot_sentiment}). Similar results emerge: both statements of \textit{intention} and \textit{calls-to-action} show strong, positive associations with the group of positive emotions. We also see meaningful negative associations between the levels of \textit{execution} and \textit{call-to-action} and the negative group. 

We note, as also observed in the more granular approach, that there are no associations at the \textit{problem–solution} level. One possible reason for this result is that this category combines quite different rhetorical stances: the identification of problems and the attribution of blame, which are logically aligned with negative emotions, and the articulation of solutions, which is more likely associated with positive emotions. Some examples of \textit{problem-solution} expressions pointing towards this can be found in Table~\ref{tab:probsol}.

\section{Validation}
To assess the performance of the classification models, three of the authors independently evaluated random subsets of the data. For the binary collective action classification task, we annotated $200$ sentences. For the multi-class collective action task, we sampled $300$ sentences proportionally to the overall class distribution. In addition, we annotated $200$ sentences for the race classification task. The detailed annotation guidelines are provided in Appendix~\ref{app:codebook}.

Table~\ref{tab:eval} reports validation results for the classifiers employed in our study to detect participation in BLM: binary expression of collective action, multi-class expression of collective action and binary expression of racial justice.

For the binary collective action task, inter-annotator agreement was relatively low (Krippendorff $\alpha=0.32$), reflecting the inherent difficulty of the task. Nonetheless, the classifier achieves a reasonably strong performance on podcast transcripts (macro F1 = $0.60$). This is comparable to its performance on Reddit comments (macro F1 = $0.65$) reported in prior work~\cite{pera_extracting_2025}, suggesting that the model generalizes well to transcribed spoken language.

In contrast, the multi-class collective action task proves substantially more challenging. Both inter-annotator agreement and classifier performance are low ($\alpha=0.40$; macro F1 = $0.22$ on podcasts, compared to $0.52$ on Reddit data). This underscores the added complexity of fine-grained classification in spoken discourse, where even human coders struggle to reach consensus. While the resulting classifier performance is modest, we view the analysis in the current work as an exploratory step that highlights the limits of current approaches and the need for further refinement when applying multi-class collective action frameworks to spoken content.

Finally, the binary racial justice task is comparatively straightforward, with higher inter-annotator agreement ($0.77$) and strong classifier performance (F1 = $0.77$).

\begin{table}[t]
\centering
\begin{tabular}{@{}lccc@{}}
\toprule
\textbf{Task} & \textbf{Krippendorff $\alpha$} & \textbf{Accuracy} & \textbf{F1} \\ \midrule
\textbf{Binary CA}        & 0.3152 & 0.6600 & 0.6047 \\ 
\textbf{Multi-class CA}   & 0.4033 & 0.2267 & 0.2169 \\ 
\textbf{Racial justice}   & 0.7737 & 0.8100 & 0.7654 \\ 
\bottomrule
\end{tabular}
\caption{Evaluation metrics. CA denotes collective action.}
\label{tab:eval}
\end{table}

\section{Discussion and Conclusion}
In this study, we provide the first large-scale, systematic analysis of activism expressed in spoken language, focusing on racial justice podcasts. We examine how expressions of collective action intersect with emotional language, a perspective that has been understudied compared to text- and image-based media.

We showed that podcast discussions about race increase in volume following BLM events (\textbf{RQ1}) and that most of such discussions express participation in collective action, with the majority of participation statements following problem identification and solution proposals (\textbf{RQ2}). 
Finally, our analysis revealed links between positive emotions and stages of collective action: \textit{caring} aligns with \textit{calls-to-action}, \textit{optimism} with \textit{intention}, and \textit{joy} with \textit{execution} (\textbf{RQ3}). 

Methodologically, this work explores how computational models can capture activism in dialogical, reflective media like podcasts; substantively, it highlights the role of constructive emotions in sustaining participation.
We found broad consistency with prior studies of interviews, protest art, and tweets from the same period, which we summarize below.

\subsection{Comparison with Existing Studies: Emotions in BLM}
Prior research on protest emotions emphasizes their temporal and functional dynamics: anger, guilt, shame, and regret often precede mobilization; hope, enthusiasm, and solidarity emerge during protest; and disappointment or frustration may follow~\cite{van_troost_emotions_2013}. Emotions thus serve motivational and communicative functions, shaping participation, group cohesion, and persuasion. In BLM, movement slogans and messaging span a wide spectrum of affect, with love and joy fostering resilience and leaders strategically guiding emotion for moral influence~\cite{hinderliter_more_2021}.

Much of the literature highlights anger as the primary mobilizing force in BLM and other movements~\cite{van_zomeren_put_2004, jasper_emotions_2018, stanley_anger_2021, okpala_analyzing_2025}, in line with broader emphasis on polarization and negative emotions~\cite{goodwin_emotions_2006}. Yet several studies complicate this picture. \citet{field_analysis_2022} identify high levels of positivity, alongside anger and disgust, in the aftermath of George Floyd’s murder, while \citet{nguyen_progress_2021} document a temporary decline in negativity during early June 2020, replaced by calls-for-action and positively framed conversations. Our results align more closely with this latter strand: we find no positive association between \textit{anger} (or other negative emotions such as \textit{sadness} and \textit{disgust}) and collective action. This divergence from the dominant view likely reflects both methodological and contextual factors: Twitter studies often capture broad hashtag conversations, whereas our focus isolates explicit calls to collective action, and podcasts tend to be more reflective than immediate or confrontational.

By contrast, positive emotions emerge in our data as consistent drivers of participation. \textit{Joy} is strongly tied to execution-level activity, echoing findings from graffiti analyses that emphasize hopeful and solidaristic frames alongside anger~\cite{cappelli_black_2020}. Interviews with BLM demonstrators in Norway likewise found that hope, paired with anger, motivated initial participation, while pride, fatigue, and agency sustained engagement~\cite{ellefsen_black_2022}. U.S.-based interviews identified a similar mix of rage, fear, hope, love, and sadness~\cite{chevrette_humanizing_2025}, while protest art analyses in Washington, D.C. revealed that anger and pride dominated written slogans but grief was more powerfully expressed in visual memorials~\cite{vance_black_2022}. Compared with these media, podcast \textit{calls-to-action} are strikingly dialogical and forward-looking, often framed in \textit{optimism} rather than confrontation. \textit{Optimism}, in turn, is strongly linked to \textit{intention}-level engagement, consistent with patterns in protest interviews~\cite{ellefsen_black_2022, chevrette_humanizing_2025} and online discussions of racial justice~\cite{nguyen_progress_2021}.

Taken together, these studies highlight the diversity and dynamism of emotions in BLM across time, media, and method. Computational analyses often foreground anger and negativity, while qualitative approaches reveal a coexistence of hope, pride, solidarity, and love with grief and rage — emotions that enable resilience and sustain mobilization. Our analysis extends this broader perspective by linking emotions to distinct phases of protest engagement. Whereas \textit{anger} and other negative emotions appear highly context- and medium-dependent, positive, forward-looking emotions (\textit{optimism}, \textit{joy}, and \textit{caring}) consistently drive collective action across podcasts, interviews, and visual protest media, suggesting that they may be central to mobilization in more reflective settings.

\subsection{Limitations and Future Work}

Despite the insights gained through our study, several limitations point to opportunities for methodological improvement.

A first limitation concerns speaker attribution, as only a subset of transcripts contains reliable labels. This constrains the analysis of conversational dynamics, which could otherwise open a valuable line of investigation~\cite{moldovan_exploring_2025}. Relatedly, relying on automatic transcriptions introduces potential classification errors. Incorporating audio features such as tone or pitch could improve accuracy in emotion and participation detection, though this would require substantial computational resources and novel methods.

Another challenge lies in classifier performance. While binary detection of participation in collective action is reasonable, performance drops markedly for the multi-class task. This reflects both the inherent difficulty of the problem and the limited generalizability of Reddit-trained models to podcast data, underscoring the need for further methodological development.

A third set of limitations relates to the study of emotions in podcasts. Classification from text alone remains particularly challenging, highlighting the importance of multimodal approaches. Moreover, as with prior work on tweets~\cite{field_analysis_2022}, our approach cannot identify the targets of emotions (e.g., protesters’ anger versus anger at protesters), limiting our understanding of how pro- and anti-groups express collective action. Positive and ambiguous emotions are also understudied, partly because existing classifiers focus disproportionately on negative emotions. Emotions such as pride, grief, and realization, which are well established in qualitative research~\cite{vance_black_2022, chevrette_humanizing_2025, nguyen_progress_2021, ellefsen_black_2022}, were omitted here due to limited classifier performance. Future work should therefore expand the range of emotions analyzed and move ``beyond Black pain to incorporate Black joy''~\cite{dunbar_black_2022}.

Beyond methodological challenges, our dataset is predominantly US- and English-centric, which limits the generalizability of findings to other contexts. Expanding analysis to international movements and cross-cultural variations would strengthen our understanding of collective action globally.

Finally, we acknowledge the ethical risks of this work. The classifiers used and the insights derived from them could be misapplied — for instance, by opponents of racial justice movements or in other political contexts. It is therefore crucial that future applications be accompanied by careful ethical reflection and safeguards against harmful use.

\section{Related Work}

\subsection{Podcasts as a Medium for Political and Social Discourse}
Podcasts have proven to be a diverse, influential, and increasingly popular medium. In 2023, 42\% of Americans aged 12 and older reported listening to podcasts in the past month, with many tuning in multiple times per week~\citep{pew_research_center_audio_2023}. The impact of podcasts extends beyond entertainment. Listeners often report significant behavioral changes as a result of their consumption, such as modifying their media preferences, altering their lifestyle, purchasing products, or even supporting political causes~\citep{shearer_black_2023}. Furthermore, the majority of listeners trust the information they hear on these shows, despite researchers' growing concerns about the spread of misinformation through this medium~\citep{brandt_echoes_2023, wirtschafter_audible_2023, wirtschafter_challenge_2021}. 

Podcasts also intersect with political participation. Recent research shows that consumption of non-mainstream political podcasts is positively associated with both believing and sharing misinformation, which in turn increases the likelihood of engaging in contentious political participation~\citep{rasul_podcasts_2025}.
Nevertheless, scholarship has largely emphasized the medium's risks, particularly its potential to foster polarization, echo chambers, and the spread of misinformation, while paying less attention to its role in mobilization for social change. This study addresses the existing gap by examining how racial justice podcasts articulate expression of collective action.

\subsection{The Black Lives Matter Movement in Digital Contexts}
Racial justice has been examined in the context of multiple media, however the majority of empirical studies rely on tweets~\cite{okpala_analyzing_2025, field_analysis_2022, nguyen_progress_2021, ince_social_2017} and focus on the divergence between \#BlackLivesMatter protests and \#AllLivesMatter
counter-protests~\cite{carney_all_2016, gallagher_divergent_2018, knierim_divergent_2024}. A notable study is that of~\citet{brown_how_2022}, who
analyzed how pro- and anti-BLM groups justified and mobilized collective action
online, finding that groups supporting the movement emphasized out-group actions while their counterparts focused on in-group identity. The reverse pattern occurred in the justification of actions, highlighting the importance of ideology and socio-structural position in online mobilization. 

~\citet{litterer_mapping_2025} showed that podcast mentions of George Floyd peaked within ten days of his murder and gradually declined. While their work demonstrates that podcasts capture timely responses to major collective events, it does not analyze specific statements of collective action nor the emotional dynamics of these conversations. Thus, although prior research has mapped the presence of BLM in digital media broadly and in podcasts specifically, we lack a systematic account of how podcast discourse constructs mobilization towards participation in collective action. Our study addresses this gap by examining podcasts as a site of collective action discourse.

\subsection{Detecting Collective Action and Emotions in Digital Traces}
Collective action theories~\cite{van_zomeren_toward_2008, furlong_social_2021, thomas_mobilise_2022} have often been operationalized to detect collective action markers from text. Some studies rely on dictionaries that map predefined terms to theoretical concepts~\cite{gulliver_assessing_2021, brown_how_2022}, while others augment these approaches with network-based models to capture relationships between terms~\cite{erseghe_projection_2023}.~\citet{smith_after_2018} introduced a lexicon of collective action to quantify the proportion of such terms within text. Despite their usefulness, lexicon-based approaches overlook context-dependent language, and fail to adapt to linguistic variation.
One of the notable developments in this area is the topic-agnostic classifier proposed by~\citet{pera_extracting_2025} and used in this study, designed to identify different levels of participation in social movements from social media posts. 

Recent work on emotion detection in text has moved beyond simple taxonomies~\cite{ekman_argument_1992, plutchik_psychoevolutionary_1982}, toward fine-grained and domain-adapted approaches. Central to this shift is the GoEmotions dataset~\cite{demszky_goemotions_2020}, which provides 58k Reddit comments annotated with 27 emotion categories, covering positive, negative, and ambiguous sentiment. Its adaptability has been demonstrated across domains~\cite{greco_e2mocase_2024}, taxonomies~\cite{field_analysis_2022}, and model architectures~\cite{liu_emollms_2024}.

Few studies examine the interaction between collective action and emotions within a single analytical framework. Our work addresses these gaps by applying state-of-the-art classifiers to podcast transcripts, allowing us to capture both expressions of participation in activism and the emotional context in discussions of racial justice. This way, we can analyze not only how participation is expressed, but also how it is emotionally framed, providing a richer understanding of racial justice discourse in digital contexts.

\bibliography{references.bib, other_refs.bib}

\clearpage 

\appendix

\setcounter{figure}{0}
\setcounter{table}{0}
\setcounter{equation}{0}
\renewcommand{\thefigure}{A\arabic{figure}}
\renewcommand{\thetable}{A\arabic{table}}
\renewcommand{\theequation}{A\arabic{equation}}

\setcounter{secnumdepth}{2}

\section{Data and Models} \label{app:datamodels}

\subsection{Licenses} \label{app:licenses}

The Structured Podcast Research Corpus is freely available at \url{https://huggingface.co/datasets/blitt/SPoRC}. The \verb|RoBERTa| model used for the collective action binary classification task is publicly available at \url{https://huggingface.co/ariannap22/collectiveaction_roberta_simplified_synthetic_weights} and the Supervised Fine-Tuning approach to the multi-class classification task can be found at \url{https://huggingface.co/ariannap22/collectiveaction_sft_annotated_only_v6_prompt_v6_p100_synthetic_balanced_more_layered}. The \texttt{Llama-3.3-70B-Instruct} model employed for the zero-shot approach is an open-source LLM released under a commercial use license\footnote{\url{https://github.com/meta-llama/llama-models/blob/main/models/llama3_3/LICENSE}}. We employ the \texttt{unsloth/Llama-3.3-70B-Instruct-bnb-4bit} version available at \url{https://huggingface.co/unsloth/Llama-3.3-70B-Instruct-bnb-4bit} of the model for more efficient memory use. The \verb|RoBERTa| classifiers for all emotions are available at \url{https://huggingface.co/SamLowe/roberta-base-go_emotions}.

\subsection{Keywords} \label{app:keywords}

\begin{table}[ht!]
\centering
\begin{tabular}{l}
\hline
\textbf{Keyword} \\
\hline
racism \\
racial justice \\
police brutality \\
police violence \\
black people \\
black person \\
george floyd \\
breonna taylor \\
trayvon martin \\
ahmaud arbery \\
sandra bland \\
tamir rice \\
eric garner \\
black lives matter \\
systemic racism \\
structural racism \\
racial inequality \\
racial profiling \\
racial bias \\
institutional racism \\
mass incarceration \\
criminal justice reform \\
abolish the police \\
defund the police \\
racial uprising \\
white supremacy \\
\hline
\end{tabular}
\caption{List of keywords used to filter episodes.}
\label{tab:keywords}
\end{table}

Table~\ref{tab:keywords} contains the keywords used in the search on all transcripts provided in the Structured Podcast Research Corpus. The list was used to filter episodes that include conversations on the topic of race.

\subsection{Computational Setup} \label{app:comp}

A combination of NVIDIA A40 and A100 GPUs on a shared HPC cluster was used for the collective action, race, and emotion classification tasks. Jobs consisting of $1,000$ episodes each were run in parallel for each task.
The binary collective action classifier was applied to all $83$M sentences contained in the $155,784$ episodes identified through the keyword search. This inference step completed in under $24$ hours.  
The subsequent multi-class collective action and racial justice classifiers were applied on the $2$M sentences identified as expressions of collective action in the first step. The multi-class collective action run took approximately one week, whereas the racial justice run required roughly $2.5$ weeks. 
The emotion classifiers were applied to all $76$M sentences contained in the final sample of $139,200$ episodes. This task was completed within a day.

\subsection{GoEmotions Classifiers} \label{app:goemo}

The GoEmotions model was trained for $3$ epochs with a learning rate of $2e-5$ and weight decay of $0.01$. We use $0.5$ to binarize the scores. Table~\ref{tab:big_emotions} contains details on all emotions in the dataset.

\section{Results} \label{app:results}
\renewcommand{\thetable}{B\arabic{table}}
\setcounter{table}{0}

\subsection{Episode Categories} \label{app:categories}

\begin{table}[ht!]
\centering
\begin{tabular}{@{}lc@{}}
\toprule
\textbf{Category} & \textbf{Proportion of episodes (\%)} \\ 
\midrule
Society       & 18.63 \\
Religion      & 15.36 \\
News          & 15.08 \\
Comedy        & 8.79 \\
Sports        & 7.93 \\
Education     & 7.08 \\
Business      & 5.23 \\
Arts          & 4.27 \\
Health        & 4.00 \\
TV            & 3.90 \\
Leisure       & 2.83 \\
Music         & 2.34 \\
Science       & 0.84 \\
Kids          & 0.83 \\
History       & 0.76 \\
Technology    & 0.71 \\
Government    & 0.68 \\
Fiction       & 0.33 \\
True Crime    & 0.28 \\
Games         & 0.09 \\
Politics      & 0.01 \\
Fantasy       & 0.01 \\
Christianity  & 0.01 \\
Entertainment & - \\
Commentary    & - \\
Natural       & - \\
Documentary   & - \\
Travel        & - \\
Running       & - \\
\bottomrule
\end{tabular}
\caption{Podcast categories of episodes with expressions of collective action related to BLM.}
\label{tab:categories}
\end{table}

Table~\ref{tab:categories} shows the proportion of episodes with expression of participation in racial justice belonging to each self-selected high-level category. Categories without proportions represent less than $0.01\%$ of the sample (14 episodes).

\renewcommand{\thetable}{A\arabic{table}}
\setcounter{table}{1}

\setlength{\tabcolsep}{4pt} 
\begin{table*}[ht!]
\centering
\begin{tabular}{@{}lp{2.2cm}p{7cm}c@{}}
\toprule
\textbf{Polarity} & \textbf{Emotion} & \textbf{Definition} & \textbf{F1} \\ 
\midrule
\parbox[t]{2mm}{\multirow{14}{*}{\rotatebox[origin=c]{90}{Negative}}} 
  & \textbf{Anger}   & A strong feeling of displeasure or antagonism. & 0.479 \\
  & \textbf{Annoyance}   & Mild anger, irritation. & 0.238 \\
  & \textbf{Disappointment}   & Sadness or displeasure caused by the nonfulfillment of one’s hopes or expectations. & 0.302 \\
  & \textbf{Disapproval}   & Having or expressing an unfavorable opinion. & 0.379 \\
  & \textbf{Disgust} & Revulsion or strong disapproval aroused by something unpleasant or offensive. & 0.453 \\
  & \textbf{Embarrassment}   & Self-consciousness, shame, or awkwardness. & 0.367 \\
  & \textbf{Fear}    & Being afraid or worried. & 0.671 \\
  & \textbf{Grief}   & Intense sorrow, especially caused by someone’s death. & 0.000 \\
  & \textbf{Nervousness}   & Apprehension, worry, anxiety. & 0.214 \\
  & \textbf{Remorse}   & Regret or guilty feeling. & 0.636 \\
  & \textbf{Sadness} & Emotional pain, sorrow. & 0.550 \\
\midrule
\parbox[t]{2mm}{\multirow{18}{*}{\rotatebox[origin=c]{90}{Positive}}}
  & \textbf{Admiration}   & Finding something impressive or worthy of respect. & 0.699 \\
  & \textbf{Amusement}   & Finding something funny or being entertained. & 0.829 \\
  & \textbf{Approval}   & Having or expressing a favorable opinion. & 0.404 \\
  & \textbf{Caring}  & Displaying kindness and concern for others. & 0.372  \\
  & \textbf{Desire}   & A strong feeling of wanting something or wishing for something to happen. & 0.496 \\
  & \textbf{Excitement}   & Feeling of great enthusiasm and eagerness. & 0.435 \\
  & \textbf{Gratitude}   & A feeling of thankfulness and appreciation. & 0.919 \\
  & \textbf{Joy}     & A feeling of pleasure and happiness. & 0.600  \\
  & \textbf{Love}    & A strong positive emotion of regard and affection. & 0.802  \\
  & \textbf{Optimism}& Hopefulness and confidence about the future or the success of something. & 0.481 \\
  & \textbf{Pride}   & Pleasure or satisfaction due to ones own achievements or the achievements of those with whom one is closely associated. & 0.000 \\
  & \textbf{Relief}   & Reassurance and relaxation following release from anxiety or distress. & 0.000 \\
\midrule
\parbox[t]{2mm}{\multirow{5}{*}{\rotatebox[origin=c]{90}{Ambiguous}}}
  & \textbf{Confusion} & Lack of understanding, uncertainty. & 0.463 \\
  & \textbf{Curiosity} & A strong desire to know or learn something. & 0.428 \\
  & \textbf{Realization} & Becoming aware of something. & 0.220 \\
  & \textbf{Surprise} & Feeling astonished, startled by something unexpected. & 0.525 \\
\bottomrule
\end{tabular}
\caption{Definitions of GoEmotions and classifier metrics.}
\label{tab:big_emotions}
\end{table*}

\renewcommand{\thetable}{B\arabic{table}}
\setcounter{table}{1}

\subsection{Emotion Distributions} \label{app:emotions}

\begin{table*}[ht]
\centering
\begin{tabular}{lc}
\hline
\textbf{Emotion} & \textbf{Percentage (\%)} \\
\hline
anger     & 1.037 \\
disgust   & 0.195 \\
fear      & 0.359 \\
sadness   & 0.886 \\
joy       & 0.795 \\
optimism  & 0.788 \\
love      & 1.402 \\
caring    & 0.844 \\
\hline
\end{tabular}
\caption{Distribution of data by emotion class.}
\label{tab:emodist}
\end{table*}

We report the distribution of the entire dataset by emotion class in Table~\ref{tab:emodist}. The percentages indicate the share of all sentences in the dataset where each emotion was detected.

\subsection{Odds Ratios} \label{app:oddsdeets}

\begin{table*}[t]
\centering
\begin{tabular}{lcccc}
\hline
\textbf{Category} & \textbf{Problem-solution} & \textbf{Call-to-action} & \textbf{Intention} & \textbf{Execution} \\
\hline
anger     & 1.06 [1.05, 1.07] & 0.64 [0.63, 0.66] & 0.31 [0.30, 0.31] & 0.21 [0.20, 0.22] \\
disgust   & 0.86 [0.85, 0.87] & 0.42 [0.40, 0.43] & 0.45 [0.43, 0.47] & 0.35 [0.33, 0.37] \\
fear      & 1.06 [1.05, 1.08] & 0.34 [0.33, 0.36] & 0.66 [0.65, 0.68] & 0.35 [0.33, 0.37] \\
sadness   & 0.93 [0.92, 0.94] & 0.36 [0.35, 0.37] & 0.58 [0.57, 0.59] & 0.44 [0.43, 0.46] \\
joy       & 0.57 [0.56, 0.57] & 0.88 [0.87, 0.90] & 1.20 [1.19, 1.22] & 2.01 [1.98, 2.04] \\
optimism  & 1.27 [1.26, 1.27] & 1.98 [1.96, 1.99] & 3.84 [3.82, 3.87] & 0.85 [0.84, 0.86] \\
love      & 0.97 [0.96, 0.97] & 1.35 [1.34, 1.37] & 2.01 [1.99, 2.03] & 0.73 [0.72, 0.75] \\
caring    & 1.33 [1.32, 1.34] & 3.15 [3.13, 3.17] & 2.01 [2.00, 2.03] & 0.95 [0.94, 0.96] \\
\midrule
negative  & 0.98 [0.97, 0.99] & 0.44 [0.43, 0.45] & 0.50 [0.49, 0.51] & 0.34 [0.32, 0.35] \\
positive  & 1.03 [1.03, 1.04] & 1.84 [1.82, 1.86] & 2.27 [2.25, 2.29] & 1.14 [1.12, 1.15] \\
\hline
\end{tabular}
\caption{Odds ratios by sentiment and emotion.}
\label{tab:oddsdeets}
\end{table*}

Odds ratios displayed in Figure~\ref{fig:lolliplot_emotions} and Figure~\ref{fig:lolliplot_sentiment} are included in Table~\ref{tab:oddsdeets}, alongside $95\%$ confidence intervals.

\subsection{Problem-solution Statements} \label{app:examples}

\begin{table*}[ht]
\centering
\begin{tabular}{lp{10cm}}
\hline
\textbf{Emotion detected} & \textbf{Problem-solution statement} \\
\hline
Anger & You know, anger at sort of what it took to create this awakening, anger at the militarization of the police force and how we're seeing them come down on peaceful protesters, anger at folks who are trying to sort of co-opt this people's movement by being agitators and starting things.\\
Disgust & But we did come prepared because prior to that day, every single day, peaceful protesters had been tear gas shot with rubber bullets, disgusting things done by our own government to peaceful protesters.\\
Fear & Cops think they have all the power in the world and I have felt it personally myself as a Hispanic man, the fear of being stopped by a police officer of any kind.\\
Sadness & It's just really sad because you don't really know, like, I don't know whether to tell my daughter to trust the police or to fear the police.\\
Caring & And the solution is that you have to communicate outside your box, you know, take the time, experience some new cultures, you know, don't let the illusion of fear overtake your thoughts and overtake your mind.\\
Joy & But I'll be, I'll be happy if anybody actually faces any justice.\\
Love & I mean, all the divides we have, we've got to love the one that we have a hard time understanding or comprehending or even respecting.\\
Optimism & And that is what gives me even more hope is that this issue has now become a global issue where minorities, in those countries are also minorities, but you see a lot of people in Amsterdam, which is a majority white country protesting for us.\\
\hline
\end{tabular}
\caption{Some examples of problem-solution statements by detected emotion.}
\label{tab:probsol}
\end{table*}

We present examples of sentences classified as \textit{problem-solution} expressions for each emotion under analysis in Table~\ref{tab:probsol}.

\section{Annotation Codebooks} \label{app:codebook}

\subsection{Collective Action}

For the codebook used to evaluate the collective action classifiers~\cite{pera_extracting_2025}, see \texttt{annotation\_codebook.pdf} in the GitHub repo \url{https://github.com/ariannap13/extract_collective_action}.

\subsection{Racial Justice}

Given the sentence from the podcast episode, classify whether the text mentions any of the following topics: race, ethnicity, systemic racism, racial justice, or police brutality. Select the appropriate label based on the content of the sentence.

Labels:
\begin{itemize}
    \item 1: The sentence mentions race, ethnicity, systemic racism, racial justice, or police brutality in any way. This could include discussions on racial disparities, experiences related to racism, references to systemic issues, advocacy for racial justice, or any mention of police brutality.
    \item 0: The sentence does not mention any of the above topics. If the sentence is unrelated to race, ethnicity, systemic racism, racial justice, or police brutality, label it as ``0.''
\end{itemize}

Guidelines for classification:
\begin{itemize}
    \item Race or ethnicity: This includes any references to racial or ethnic groups, discussions of cultural identity, or distinctions between races or ethnicities. Example: ``The challenges faced by Black communities are often overlooked.''
    \item Systemic racism: The sentence addresses the existence or impact of racism as a systemic issue that affects institutions, policies, or societal structures. Example: ``The criminal justice system has been criticized for perpetuating racial inequalities.''
    \item Racial justice: The sentence talks about efforts, movements, or the need for justice related to racial equality, or reform to address racial inequities. Example: ``Activists are pushing for policies that promote racial justice.''
    \item Police brutality: The sentence refers to or discusses the use of excessive force by law enforcement, particularly against people of color. Example: ``Protests erupted following another incident of police brutality.''
\end{itemize}

\section{Prompts} \label{app:prompts}

Prompts for the multi-class collective action classifier are available in the appendices of~\citet{pera_extracting_2025}. The prompt used in the zero-shot racial justice classifier is below.

\begin{promptbox}{LLM Racial Justice Task}
Classify whether the text mentions race, ethnicity, systematic racism, racial justice, or police brutality in any way (``1'') or not (``0'').
A sentence is considered to express racial justice if it fits in any of the following descriptions:
* The sentence includes any references to racial or ethnic groups, discussions of cultural identity, or distinctions between races or ethnicities.
* The sentence addresses the existence or impact of racism as a systemic issue that affects institutions, policies, or societal structures.
* The sentence refers to or discusses the use of excessive force by law enforcement, particularly against people of color. 
* The commenter is describing their personal experience taking direct actions towards a common goal.
Return the label ``1'' or ``0'' based on the classification.
Comment: {text}
Label: 
\end{promptbox}

\end{document}